\documentclass[runningheads]{llncs}
\usepackage[T1]{fontenc}
\usepackage{orcidlink}
\usepackage[misc]{ifsym}
\usepackage{subfiles}
\usepackage{makecell}
\usepackage{enumitem}
\usepackage{multirow}
\usepackage{longtable}
\usepackage{graphicx}
\usepackage[dvipsnames]{xcolor}
\usepackage{tabularx}
\usepackage{soul}
\usepackage{float}
\usepackage{pifont}
\usepackage{subfiles}

\newif\ifredact
\redactfalse  

\newcommand{\tsk}{\textsc{TalkSketch}}

\newcommand{\dquote}[1]{\textit{``#1''}}

\begin{document}
\title{TalkSketch: Multimodal Generative AI for Real-time Sketch Ideation with Speech}
\titlerunning{TalkSketch}
%
\author{
Weiyan Shi\inst{1}\orcidlink{0009-0001-6035-9678} \and
Sunaya Upadhyay\inst{2}\orcidlink{0009-0004-9613-7593} \and
Geraldine Quek\inst{1}\orcidlink{0000-0003-2864-3860} \and
Kenny Tsu Wei Choo\inst{1}\orcidlink{0000-0003-3845-9143}\textsuperscript{(\Letter)}
}
\authorrunning{Weiyan Shi et al.}
%
\institute{
\textsuperscript{1}Singapore University of Technology and Design, Singapore \\
\email{weiyanshi6@gmail.com, geraldine\_quek@sutd.edu.sg, kennytwchoo@gmail.com} \\
\textsuperscript{2}Carnegie Mellon University, Pittsburgh, PA, United States \\
\email{sunayau@andrew.cmu.edu}
}
\maketitle              
\begin{abstract}
    Sketching is a widely used medium for generating and exploring early-stage design concepts. While generative AI (GenAI) chatbots are increasingly used for idea generation, designers often struggle to craft effective prompts and find it difficult to express evolving visual concepts through text alone. In the formative study (N=6), we examined how designers use GenAI during ideation, revealing that text-based prompting disrupts creative flow. To address these issues, we developed \tsk{}, an embedded multimodal AI sketching system that integrates freehand drawing with real-time speech input. \tsk{} aims to support a more fluid ideation process through capturing verbal descriptions during sketching and generating context-aware AI responses. Our work highlights the potential of GenAI tools to engage the design process itself rather than focusing on output.
\keywords{generative AI \and sketching \and talking \and creativity support}
\end{abstract}

\section{Introduction}

During early-stage design, sketching plays a central role as an improvisational, open-ended, and dynamic practice~\cite{landay1995interactive,suwa2022roles}. Designers frequently switch between design phases and iterate their sketches to explore alternatives and refine ideas toward promising directions~\cite{buxton2010sketching}. To enhance efficiency in this formative stage, researchers have explored integrating generative AI (GenAI)~\cite{fui2023generative} into sketch-based design workflows~\cite{davis2025sketchai,lin2025inkspire,zhang2023generative,zhou2025partstickers}. With advances in multimodal large language models (LLMs), GenAI now offers stronger creative capabilities~\cite{vinker2025sketchagent,xu2024llm,ccelen2024design}, making it increasingly suitable for supporting designers in early-stage ideation.

However, despite the growing use of GenAI chatbots in idea generation, designers often struggle to craft effective prompts and to express evolving visual concepts through text alone. This command-based interaction paradigm~\cite{subramonyam2024bridging} places the burden on designers to direct the system, overlooking other forms of input that may be more natural and intuitive during creative work~\cite{kim2009study,purcell1998drawings}. In practice, designers often verbalise ideas spontaneously while sketching, yet such contextual cues are rarely captured by current systems.

To better understand these challenges, we conducted a formative study (N=6) examining how designers use existing GenAI tools for early-stage sketch ideation. Our findings revealed that text-based prompting often interrupts creative flow and creates a disconnect between ideation and sketching activities.

Based on these insights, we developed TalkSketch, a sketching interface with a multimodal AI chatbot that enables users to draw and verbalise ideas simultaneously. TalkSketch captures designers’ spoken descriptions during sketching and generates contextually relevant AI responses, aiming to support a more fluid and natural ideation process.

Our research contributes:
\begin{enumerate}
\item Findings from a formative study (N=6) revealing key challenges designers face when using current GenAI tools for early-stage sketch ideation.
\item \tsk{}: a sketching interface with a multimodal AI chatbot that enables users to draw and verbalize ideas simultaneously.
\end{enumerate}
\section{Related Work}

\subsection{Multimodal Human–AI Interaction for Creative Ideation}

Conversational User Interfaces (CUIs) enable dialogue-based interaction that mimics human conversation~\cite{mctear2002spoken} and are widely deployed in chatbots~\cite{folstad2017chatbots} and voice-activated assistants such as Amazon Alexa and Apple’s Siri~\cite{de2020intelligent}. Recent advances in Generative AI (GenAI)~\cite{feu2024generative,tank2024metacognitive}, including large language models (LLMs) such as ChatGPT~\cite{achiam2023gpt} and multimodal assistants such as Gemini~\cite{team2024gemini}, have made conversational human–AI interaction broadly accessible and flexible. These tools are increasingly applied in domains such as education~\cite{shaer2024ai}, healthcare~\cite{hedderich2024piece}, and creative work~\cite{angert2023spellburst,zhang2025breaking}, supporting users through natural language dialogue.

A central concern in these systems is whether users can express intent clearly and efficiently. Prior work has explored multimodal input and interface strategies to improve intent communication. Hu et al.~\cite{hu2025gesprompt} introduced \textit{GesPrompt}, which combines co-speech gestures with voice input to enable more natural intent expression in extended reality environments. Cho et al.~\cite{cho2025persistent} proposed \textit{Persistent Assistant}, integrating embodied input and multimodal feedback for seamless everyday interactions. Other work embeds LLMs into direct manipulation environments, mapping graphical user interface actions or visual edits into structured prompts~\cite{angert2023spellburst,masson2024directgpt}, showing that embedding language models within interactive contexts can reduce the burden of verbose prompting. Yet, most of these systems are designed for short, directive, or domain-specific tasks such as code editing or data visualisation, rather than open-ended creative ideation.

In design, creativity-support tools increasingly integrate GenAI into mainstream workflows. Commercial platforms such as Adobe Firefly~\cite{adobefirefly2025tools}, Canva Magic Studio~\cite{canva2025overview}, and Figma AI~\cite{figmaai2025update} demonstrate how generative models assist in rapid visual exploration, while research prototypes extend this trend toward more expressive input. For example, \textit{DesignPrompt}~\cite{peng2024designprompt} allows designers to compose prompts using text, colour, and imagery, and \textit{DesignWeaver}~\cite{tao2025designweaver} introduces palette-based refinement to support iterative design. Similarly, sketch-based interfaces such as \textit{Inkspire}~\cite{lin2025inkspire} and \textit{SketchAI}~\cite{davis2025sketchai} demonstrate how freehand input can guide image generation and support analogical inspiration.

Despite these developments, two key gaps remain.
First, existing multimodal CUIs focus primarily on explicit prompts and discrete commands, overlooking more spontaneous forms of intent expression such as speech or thinking aloud during creative work.
Second, current creativity-support tools rely mainly on visual and textual input, rarely considering how spoken language—produced naturally during sketching—might complement visual ideation.

Our work addresses these gaps by investigating how combining speech and sketch input can support fluid, multimodal interaction with GenAI during early-stage design ideation.

\subsection{Talking as a Novel Input Modality for Sketching}
Speech is also a natural modality for externalising thought, long recognised in systems such as \textit{SHRDLU}~\cite{winograd1972understanding} and \textit{Put-That-There}~\cite{bolt1980put} from the 1970-80s. It enables real-time, low-friction expression of ideas, making it especially valuable in early-stage design when thoughts are fluid and evolving~\cite{fry1979physics}.
Sketching and speech together can serve as two intuitive, complementary interaction methods in design and creative domains, allowing users to express nuances that one modality alone might omit ~\cite{adler2007speech}.

Recent works demonstrate how combining sketch and speech supports more natural and expressive communication: Rosenberg et al.~\cite{rosenberg2024drawtalking} introduced \textit{DrawTalking}, which combines sketching with storytelling through speech to construct interactive animated worlds, while Giunchi et al.~\cite{giunchi2021mixing} showed that integrating sketch and speech for 3D model retrieval in virtual environments helped overcome the limitations of sketch-only input. Cheng et al.~\cite{cheng2025aiawareness} also explored speech and sketch as inputs for enhancing communication with AI through context awareness in an exploratory Wizard-of-Oz study.  

While prior research highlights how speech can enhance collaboration with AI, the impact of naturally produced speech during sketching on subsequent GenAI interactions remains unexplored. Our work therefore systematically investigates sketching-while-talking as a combined input modality for early-stage design ideation. Rather than treating sketching and speech as separate channels, we integrate both within a multimodal LLM chatbot to examine whether verbal descriptions and visual strokes together can improve AI alignment with user intent.
\section{Formative Study: Understanding Early-Stage Design Workflow with GenAI}

We first conducted a formative study, combining a design task with interviews, to understand how designers integrate GenAI into early-stage design and to identify opportunities for multimodal AI support to inform our subsequent system design in Section \ref{sec:talksketch}. Our institution's ethics review board approved this study, and we obtained informed consent from all participants.

\subsubsection{Participants}
We recruited six participants (2 female, 4 male) with a basic background in design; at the very least, they had completed one design-related course (e.g., \textit{Urban Sketching} or \textit{Design Thinking and Innovation}).
The participants included three students, two entry-level designers, and one experienced design practitioner, with design backgrounds spanning architecture, furniture, interior, robotics, and electronic product design.
All participants had 2-5 years of part/full-time experience in design. 
They received approximately USD 7.8 as compensation for completing the 1-hour study.

\subsubsection{Study Protocol}
The study consists of three parts: pre-task interview (10 minutes), design task (30 minutes), and post-task interview (20 minutes).
The pre-task interview captures participants' design backgrounds, sketching habits, and prior experiences with GenAI tools.
For the design task, we asked participants to design a household bread toaster and to express as many design ideas as possible under 30 minutes using a sketching application such as Goodnotes\footnote{https://www.goodnotes.com/} or Procreate\footnote{https://procreate.com/}. 
They were permitted to utilize their personal devices--to ensure a familiar arrangement--or our Apple iPad Pro 13" and Apple Pencil set up with popular sketching apps (e.g., Goodnotes, Procreate) and GenAI tools (e.g., ChatGPT (GPT-4o)\footnote{https://chatgpt.com/}, Gemini\footnote{https://gemini.google.com/app}, Midjourney\footnote{https://midjourney.com}). 
The post-task interview then examined participants' views on the AI's role in supporting or limiting ideation, and captured their aspirations and expectations for an "ideal" AI assistant.

\subsubsection{Data Analysis}
We analysed the qualitative data by examining both participants' interactions with sketching and AI tools during the design tasks and their reflections in post-study interviews. This allowed us to capture overall usage patterns, such as how participants sketched, engaged with GenAI tools, and iterated on ideas, while also identifying the challenges and gaps they experienced with current tools. These insights inform the design goals of our system.

\subsection{Overall Usage Patterns and Challenges}

We observed three recurring usage patterns in how participants incorporated GenAI into early-stage ideation during the design task. These patterns, along with the reported challenges, surfaced during the post-interview.

\subsubsection{Pattern 1: Using GenAI for research and ideation.}
Participants often used ChatGPT to explore design problems, define target users, or to inspire new directions. For instance, P1 asked about commonly found toaster issues and focused on cleaning functions, while P3 explored different user groups to inspire persona-driven designs. These uses supported functional exploration, but initial AI responses were often \dquote{too generic} (P3) and only became actionable after repeated prompt refinement.

\subsubsection{Pattern 2: Using GenAI to render sketch-based ideas.}
Several participants uploaded their sketches to ChatGPT Image or Gemini, or relied on text prompts, to visualise concepts. For instance, P2 uploaded a sideways-eject sketch, while P5 asked Gemini for a toaster with butter and egg compartments. However, the outputs often failed to match intent. P1 described the results as \dquote{kind of crazy,} P2 concluded, \dquote{I might have to redraw the whole thing so AI can understand,} and P5 added, \dquote{It takes too much time, I'd rather just draw.} Many participants ultimately returned to manual sketching for better clarity and control.

\subsubsection{Pattern 3: Iterative loops across sketching, prompting, and referencing.}
Participants often moved back and forth between tools. P3 only used text in ChatGPT to imagine playful forms (e.g., cat- or book-shaped toasters), while P1 first researched in ChatGPT, then sketched, went back to sketching after clarifications, and finally tried rendering images. P1 noted that \dquote{uploading sketches to ChatGPT was not so easy to operate on iPad.} This reflected how tool-switching slowed the process and introduced friction, especially for those who wanted to use image generation (P1, P2, P5, P6).

\subsubsection{Challenges across patterns.}
Across these usage patterns, participants encountered three main challenges: (1) AI responses that were too generic or required extensive refinement, (2) mismatched or low-quality image outputs that failed to convey intent, and (3) fragmented workflows due to frequent switching between different tools. These issues often interrupted the ideation flow and made participants rely more heavily on manual sketching for clarity and control.

\subsection{Design Goals for \tsk{}}
During the post-interview, participants proposed several workflow improvements to enhance the usefulness and usability of AI tools in early-stage design, particularly in terms of integration, input flexibility, and contextual responsiveness.

\subsubsection{Goal 1: Integrate AI with sketching tools.}
\label{sec:goal_1}
Rather than operating AI separately, participants hoped for tighter integration of AI within their sketching environment. P5 imagined an AI assistant \dquote{embedded directly into sketching app} that could notice overlooked design opportunities or recommend new features. P6 likewise wanted the AI to interpret her rough visuals directly and \dquote{add internal or functional details} on top of her own sketches. They expressed that embedded support would reduce switching between tools and allow AI to respond more fluidly to real-time ideation.

\subsubsection{Goal 2: Reduce designers' fatigue with long prompts.}
\label{sec:goal_2}
Many participants (P2, P5, P6) expressed fatigue from repeatedly typing long prompts to clarify their intent. P2 admitted he was \dquote{too lazy to retype the whole thing,} while P5 preferred to \dquote{just draw} instead of describing every detail. Participants advocated for expanded input modalities such as annotated sketches, voice commands, or real-time drawing, as alternatives to laborious text input. P6, for instance, envisioned being able to sketch a toaster form and have the AI \dquote{fill in the mechanism} without extra explanation. Such input flexibility would better match their natural workflows and reduce prompt clarification during ideation.

\subsubsection{Goal 3: A more proactive and context-aware AI.}
\label{sec:goal_3}
Participants wanted AI to behave less like a passive tool and more like a design partner capable of understanding the user's intent. P3 proposed that the assistant \dquote{should know what I'm trying to do} P4 suggested that AI should \dquote{see what I'm drawing and just give suggestions,} and suggested a friendly avatar that feels more responsive and approachable. P2 imagined a system that could \dquote{first discuss the sketch with me} before trying to render anything, to avoid miscommunication. This desire for AI's proactivity includes the ability to respond to ambiguous sketches, partial ideas, or evolving concepts in real time, without relying solely on explicit step-by-step instructions.
\section{\tsk{}: System Design and Development}
\label{sec:talksketch}

\begin{figure*}[t]
\centering
\includegraphics[width=\textwidth]{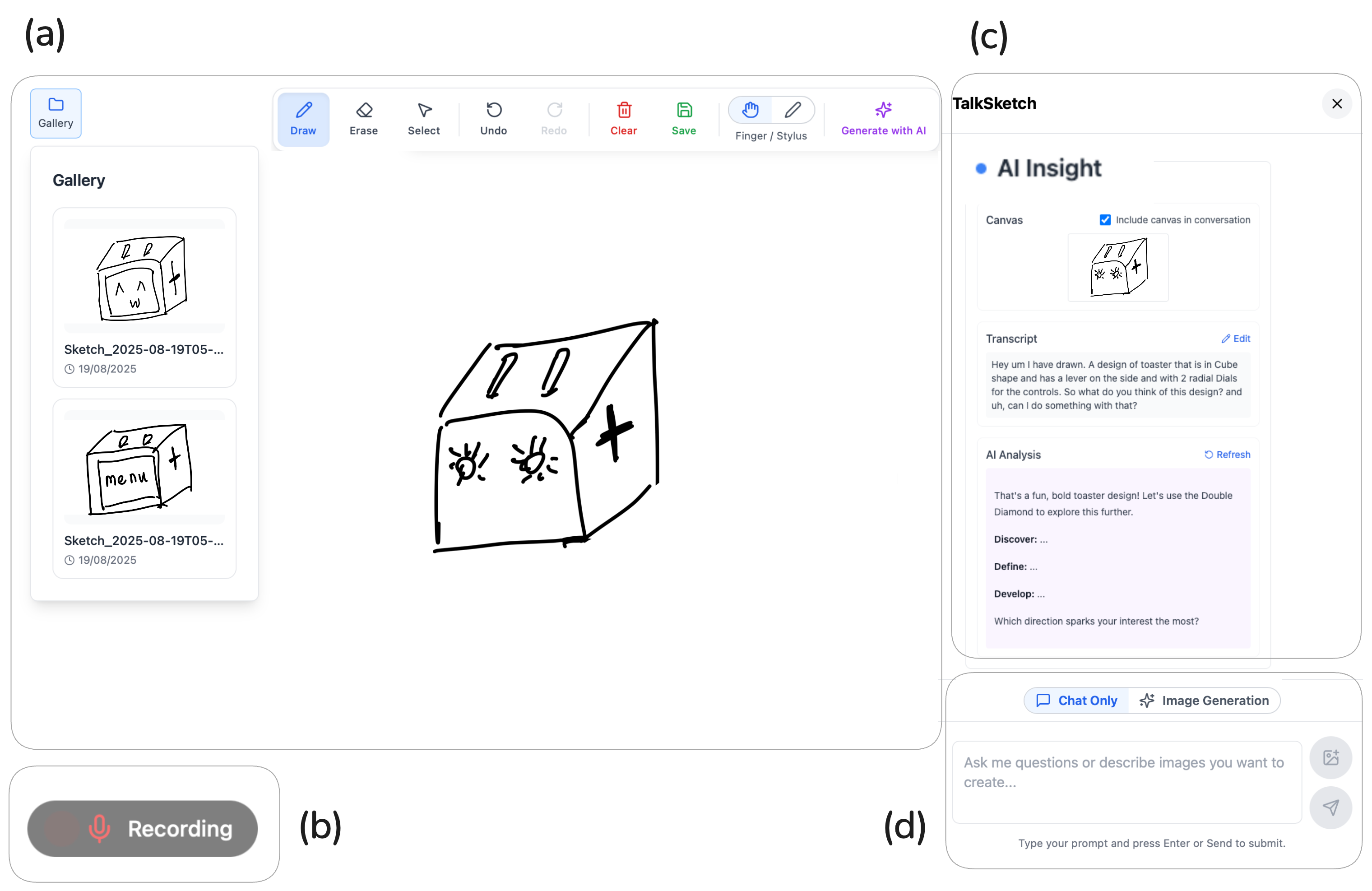}
\caption{
Overview of the \tsk{} system interface.
\textit{(a)} \textbf{Sketching} module: Users draw product concepts (e.g., a toaster) using stylus input on the canvas. The interface includes a sketch gallery, drawing toolbar, and a button to launch the \textbf{Multimodal AI Chatbot}.
\textit{(b)} \textbf{Talking} module: Voice recording captures the user's thinking aloud during sketching.  
\textit{(c)} AI Insight panel shows automatic feedback based on the sketch and spoken transcript.  
\textit{(d)} Text and image generation interface for multimodal interaction with the AI.  
Together, (c) and (d) constitute the \textbf{Multimodal AI Chatbot} module.
}
\label{fig:ui}
\end{figure*}

We first describe the technical implementation of \tsk{} (Figure~\ref{fig:ui}), which integrates three core components: a digital sketching canvas for creating and managing drawings, to address the goal of integrating GenAI into sketching tools (Sec~\ref{sec:goal_1} Goal 1); a speech capture module that transcribes the user's speech during sketching, aiming to support the goal of reducing fatigue from long prompts (Sec~\ref{sec:goal_2} Goal 2); and a multimodal AI chatbot that combines automatic insights with interactive text and image generation, aiming to make AI more proactive and context-aware (Sec~\ref{sec:goal_3} Goal 3). We then illustrate how these components work together through a system walkthrough using a concept toaster design example (Figure~\ref{fig:c3-flow}).

\subsection{Technical Implementations}
\tsk{} integrates a unified multimodal AI chatbot that aims to support both proactive and user-initiated interaction. It comprises three complementary modules: Sketching, Talking, and Multimodal AI Chatbot.

\subsubsection{Sketching Module}
The sketching canvas (Figure~\ref{fig:ui}a) serves as the central workspace where participants draw early-stage design ideas using a stylus or touch input. Built using \textit{Fabric.js}\footnote{\url{http://fabricjs.com/}}, the canvas supports drawing, erasing, selection, undo/redo, and canvas reset as illustrated in Figure \ref{fig:ui}. The top toolbar consolidates essential drawing controls along with the \textit{Generate with AI} button. To facilitate interaction between users and AI, users can select any region of the sketching canvas and export it into the chatbot as part of a multimodal prompt. The \textit{Save to Gallery} button allows users to save their canvas to the gallery once they are content with it; saved canvases can also be retrieved through the \textit{Gallery}.

\subsubsection{Talking Module}
\tsk{} augments the sketching experience with real-time voice capture to support think-aloud workflows. Audio recording initiates automatically when users are sketching (without the chatbot open), and the recording symbol will be turned on simultaneously to indicate this. Once the user opens the AI Chatbot, the recording stops and the captured audio is streamed for low-latency transcription via Google Cloud Speech-to-Text\footnote{\url{https://cloud.google.com/speech-to-text}}.

\subsubsection{Multimodal AI Chatbot Module}
The multimodal AI chatbot (see Figures~\ref{fig:ui}c and~\ref{fig:ui}d) has two main components: (1) \textit{AI Insights}, which automatically generate reflective feedback based on user sketches and verbal ideation, and (2) a \textit{Multimodal Chatbot Interface}, where users can engage the AI through both text and sketch input.
Both components share a common back-end powered by the \textit{Gemini} series of multimodal models\footnote{\url{https://ai.google.dev/gemini-api/docs/models}}
, known for their strong cross-modal reasoning capabilities and broad adoption across creative and analytical tasks~\cite{team2024gemini,yue2024mmmu}.
Specifically, the text-based conversational features are supported by Gemini 2.0 Flash, while image generation is handled by Gemini 2.5 Flash Image. Both models accept multimodal inputs such as text and images, but they serve different purposes. Gemini 2.0 Flash is a fast multimodal-reasoning variant that produces text-only output, whereas Gemini 2.5 Flash Image is an image-generation variant of the Gemini 2.5 family, specialised for producing high-quality images from multimodal prompts. The \textit{Multimodal AI Chatbot Module} maintains a unified conversation history that persists throughout the entire session and is consistently shared across both models, ensuring coherent context when switching between text and image generation.

The first feature of the Multimodal AI Chatbot, \textbf{AI Insights} (Figure~\ref{fig:ui}c), provides proactive, structured, and reflective feedback based on the user’s current sketch and verbal input. It is automatically triggered when the user clicks on \textit{Generate with AI}, requiring no explicit prompting. Each time it is activated, the module generates a new response based on the latest voice transcript and the current sketch on the canvas. The transcript is also displayed and can be edited by the user, which immediately refreshes the AI Insights response to reflect the updated input.

On the back-end, the AI Insights module is powered by \textit{Gemini~2.5~Flash} and customised to simulate the role of a design thinking expert. Following the Double Diamond framework~\cite{tschimmel2012design}, it guides users through the \textit{Discover} and \textit{Define} stages by identifying potential user needs, pain points, and framing design questions, before suggesting several exploratory directions to pursue. Two complementary prompt templates were developed to support different phases of the ideation process. When a user begins drawing on a new canvas, the system automatically triggers the \textit{Kickoff Prompt} to initiate ideation; in subsequent interactions, as the user continues sketching or revising their transcript, the \textit{Refine Prompt} is triggered to provide iterative feedback aligned with the evolving design concept. These prompts were iteratively tested to balance interpretability, creativity support, and brevity of response.

The final versions of the prompts are shown below:

\begin{quote}
\textit{\textbf{Kickoff Prompt:}}
\textit{``Act as a design thinking expert: based on the transcript and sketch canvas, identify what the user is trying to design, then—using the Double Diamond framework—guide them through Discover and Define by highlighting potential user needs, pain points, and framing questions, and finally offer 3–4 concise design directions in an encouraging and curious tone (around 100 words).''}
\end{quote}

\begin{quote}
\textit{\textbf{Refine Prompt:}}
\textit{``Act as a design thinking collaborator: based on the updated transcript and sketch canvas, briefly summarise what the user is currently designing or refining, reflect their key idea in one or two sentences, suggest 1–2 small ways to expand or clarify it, and end with 1–2 open-ended questions to help further develop the concept in a supportive, conversational tone (around 80–100 words).''}
\end{quote}

The second feature, \textbf{Multimodal AI Chat Interface} (Figure~\ref{fig:ui}d), facilitates open-ended conversations with a chatbot. Users may provide input either by typing text prompts or by utilising the iPad's integrated Voice Dictation functionality. In addition, the interface allows users to export selected regions of their sketch canvas into the chatbot as image-based inputs.

The chatbot supports two output modes: \textit{Text Generation Mode} for verbal suggestions and \textit{Image Generation Mode} for visual inspiration.
In the \textit{Text Generation Mode}, users can either submit only text input or combine their text prompt with a sketch, allowing the system to interpret both modalities but produce a text-only response.
In the \textit{Image Generation Mode}, users may similarly provide a text-only prompt or combine a text prompt with a freehand sketch, both of which will generate an image accompanied by a matching text description.
The generated images can then be imported back into the canvas to serve as visual references or be integrated into the sketch, enabling users to freely incorporate AI-generated content as they iteratively refine their designs.

\subsection{System Walkthrough}

\begin{figure*}[t]
\centering
\includegraphics[width=\linewidth]{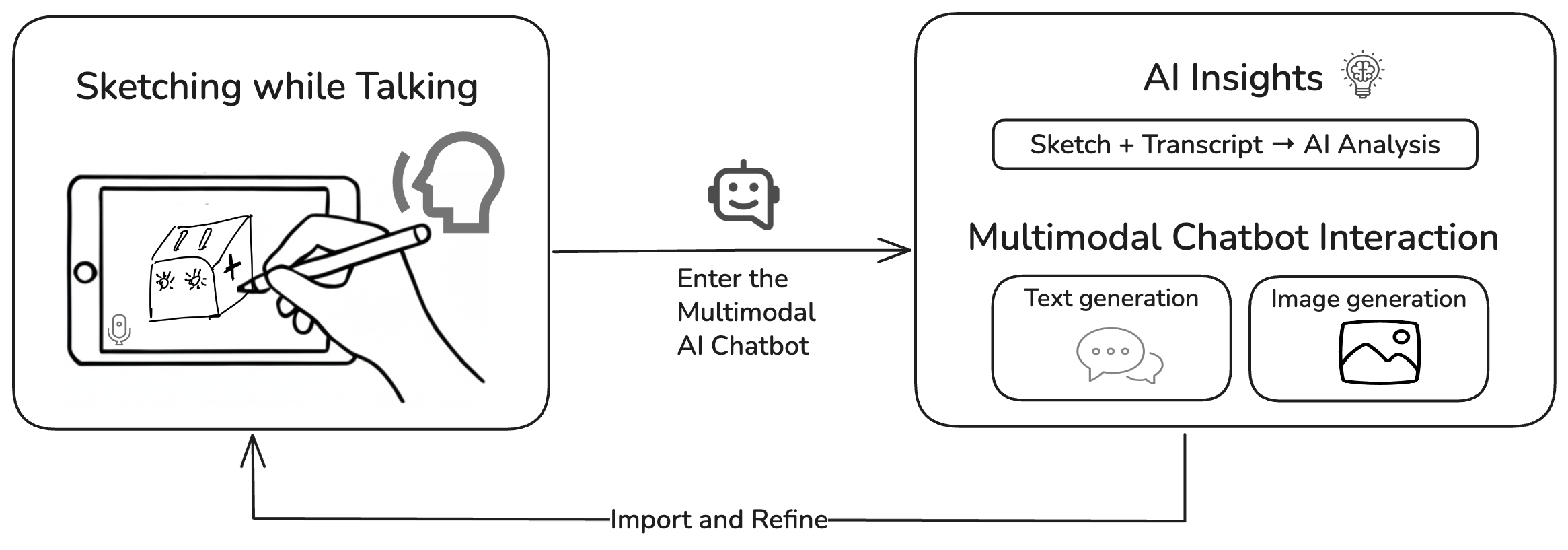}
\caption{
        The workflow of the \tsk{} system. The process starts with \textbf{Sketching with Talking}, where users draw freely on a tablet while voicing their ideas. These inputs are channelled into the \textbf{Multimodal AI Chatbot}, which generates AI Insights based on the user's sketch and transcript. Users then engage in exploration via \textbf{Multimodal Interaction}, with text or image generation to improve their design ideas. Finally, users can export AI-generated images back to the canvas for further sketching and refinement.
        }
\label{fig:c3-flow}
\end{figure*}

\tsk{} aims to support designers in early-stage ideation by combining freehand sketching, verbal thinking, and multimodal AI assistance. We illustrate its use through a walkthrough exemplar (Figure~\ref{fig:c3-flow}) where a fictional user \textit{Sky} is tasked with designing a household toaster.
Her interaction follows a natural flow: sketching while speaking aloud, triggering the chatbot interface, reviewing automatically generated AI insights, and exploring further via multimodal interaction.

\subsubsection{Sketching while Talking.}
\textit{Sky} begins the design session by drawing a rough concept of a cube-shaped toaster with symbolic radial dials and a handle (see Figure~\ref{fig:ui}a). As she sketches, she verbalises her thoughts, \dquote{I'm thinking of something bold and square, with a dial for heat control.} Her speech is recorded and transcribed in real time (see Figure~\ref{fig:ui}b), forming a synchronised verbal-visual trace of her ideation process.

\subsubsection{Entering the Chatbot and Viewing AI Insights.}
When \textit{Sky} clicks \textit{Generate with AI}, the system enters the chatbot interface (see Figure~\ref{fig:ui}d) and immediately displays an \textit{AI Insight} (see Figure~\ref{fig:ui}c). This insight is automatically generated from her sketch and think-aloud transcript, offering concise design reflections based on the Double Diamond design thinking framework.
The system suggests \emph{emphasising geometric form} or \emph{exploring tactile dials} as directions worth pursuing. Sky reads the insight and finds it helpful in framing her design direction. She realises she had forgotten to mention a few details aloud--such as her intention for the toaster to have a retractable cord--and decides to edit the transcript directly. After adding the missing information, she regenerates the \textit{AI Insight} to see how the suggestions change in response to the fuller context.

\subsubsection{Exploring via Multimodal Interaction.}
Sky then decides to follow up on the AI’s suggestions. She first types into the chatbot input box: “Could you give me some ideas for drawing a novel toaster?" In \textit{Text Generation Mode}, the chatbot responds with directions such as transparent exteriors and interactive touchscreens. Sky incorporates several of these into her sketch. To visualise the design, she switches to \textit{Image Generation Mode}, prompting: \dquote{Could you generate a realistic product based on my sketch?} The system returns a refined visual that preserves key features from her drawing.

After a few rounds, Sky finds typing increasingly tedious and instead uses the built-in voice input feature. Her verbal prompt, \dquote{What would a friendlier version look like?}, is then transcribed and submitted automatically. The system responds with a new image featuring rounded corners, soft colours, and a smiling interface. Sky finds the result inspiring and chooses to export the AI-generated image back into the canvas, using it as a visual reference while continuing to sketch and refine her final design.

\section{Potential Results}

\subsection{Enhancing Intent Expression and Interaction Naturalness}
Building on prior research in multimodal CUIs~\cite{cho2025persistent,hu2025gesprompt,masson2024directgpt}, we expect that \tsk{} will help designers communicate intent more fluidly by combining verbal and visual cues. While existing conversational AI systems require users to articulate ideas through discrete text prompts, \tsk{} allows designers to “think aloud” while drawing, preserving the spontaneity of natural dialogue~\cite{he2025enhancingintentunderstandingambiguous,lan2025contextual,a18040214}. This interaction form may reduce the cognitive effort involved in translating abstract ideas into prompts, a problem repeatedly identified in prompt-based creative tools~\cite{subramonyam2024bridging,wang2025enhancingcodellmsreinforcement}. By aligning the AI’s interpretation with both speech and sketch input, users may experience the system as a more intuitive and responsive collaborator, one that feels closer to a design partner than a command-driven assistant. We therefore anticipate higher ratings of naturalness, communication clarity, and user control, reflecting an improved sense of mutual understanding between human and AI.

\subsection{Supporting Creativity, Reflection, and Flow}
Consistent with prior work on creativity-support systems~\cite{davis2025sketchai,kang2020towards,lan2025mappo,lin2025inkspire,peng2024designprompt,tao2025designweaver}, \tsk{} is designed to support early-stage ideation rather than polished production. We expect that integrating speech and sketch will enable designers to sustain creative flow by externalising thoughts continuously instead of interrupting the process to type or reformulate text. The AI Insight mechanism may further stimulate reflection by generating contextually relevant feedback from verbal and visual cues. Designers might pause to elaborate on their thinking, reinterpret sketches, or refine concepts based on AI suggestions, which echoes the dialogic creativity patterns observed in prior sketch-based systems~\cite{buxton2010sketching,suwa2022roles}. This dynamic may lead to richer ideation traces and a stronger sense of co-evolution between human and AI ideas. Potential outcomes include higher perceived creativity support, increased exploratory behaviour, and a shift from command–response interaction to reflective conversation.
\section{Limitations}


This system has several limitations. First, because it depends on speech input, any transcription errors caused by background noise, unclear pronunciation, or the system mishearing words may lead to incorrect speech input. These mistakes may cause errors for the model to combine speech with sketches and to understand the user’s design intent. Second, in some cases, users may prefer not to speak when they are sketching alone. In such cases, the system may need to rely only on sketches and user prompting in the chatbot to support their needs, reducing a layer of speech information. However, as early-stage sketches are often incomplete or ambiguous, combining them with short or minimal prompts may make it difficult for the system to understand user needs compared to with speech information. Third, the system currently interprets speech and sketch inputs as a single combined chunk within a period of time, which inevitably leads to some information loss. Different parts of a spoken description may refer to different regions in the sketch, yet the system cannot yet distinguish these finer-grained correspondences. As a result, important links between what is said and what is drawn may be missed, limiting the precision with which user intent is understood.
\section{Conclusion and Future Work}
This paper presented \tsk{}, a multimodal generative-AI sketching system that enables designers to ideate by sketching while speaking. Through a formative study with six designers, we identified key challenges in using existing GenAI chatbots for early-stage ideation, particularly the difficulty of translating evolving visual ideas into effective text prompts. To address these issues, \tsk{} integrates freehand drawing with real-time speech input, allowing users to externalise ideas more fluidly and engage in continuous dialogue with an embedded multimodal AI chatbot. By linking verbal and visual expression, \tsk{} demonstrates how conversational multimodal interfaces can support more natural and reflective design workflows.

Future work will involve a controlled user study to systematically evaluate how \tsk{} affects the naturalness of human–AI interaction and the perceived creativity of the design process. This next step aims to provide empirical evidence for the benefits of sketch-and-speech interaction in multimodal generative design tools. Beyond creativity support, it will also be valuable to explore how this interaction style can generalise to other settings, such as live demonstrations where rapid idea communication is essential, online classrooms where instructors sketch while explaining concepts, and collaborative design reviews where teams annotate evolving visuals. Examining these broader use cases may reveal additional opportunities for applying sketch-and-speech interaction as a more versatile interface paradigm.

\bibliographystyle{splncs04}
\bibliography{main}





\end{document}